\documentstyle[12pt,amssymb,epsf]{article}
\title{ 
Asymmetric particle systems on $\Bbb{R}$}
\author{J. Krug$^1$ and J. Garc\'{\i}a$^2$ \\
{\small 1.Fachbereich Physik, Universit\"at GH Essen, D-45117 Essen, Germany} \\
{\small 2. Instituto de Matem\'atica e
Estatistica, Universidade de S\~{a}o Paulo}  \\
{\small Cx. Postal 66281, 
05315-970 S\~{a}o Paulo, Brazil}
}

\date{\today}

\begin{document}

\maketitle
\begin{abstract}
We study interacting particle systems on the real line
which generalize the Hammersley process [D. Aldous and 
P. Diaconis, Prob. Theory Relat. Fields {\bf 103}, 199-213 (1995)].
Particles jump to the right to a randomly chosen point between
their previous position and that of the forward neighbor,
at a rate which may depend on the distance to 
the neighbor. A class of models is identified for
which the invariant particle distribution is Poisson. 
The bulk of the paper is devoted to a model
where the jump rate is constant and the 
jump length is a random fraction $r$ of the distance to the
forward neighbor, drawn from a probability density
$\phi(r)$ on the unit interval. This is  
a special case of the random average process of Ferrari and
Fontes [El. J. Prob. {\b 3}, Paper no. 6 (1998)]. The discrete
time version of the model has been considered previously
in the context of force propagation in granular media
[S.N. Coppersmith {\em et al.}, Phys. Rev. E {\bf 53}, 4673 (1996)].
We show that the stationary two-point function of particle spacings 
factorizes for any choice of $\phi(r)$.
Under the assumption that this implies pairwise independence, the
invariant density of interparticle
spacings for the case of uniform $\phi(r)$
is found to be a gamma distribution with parameter
$\nu$, where $\nu = 1/2$, 1 and 2 for continuous time, backward
sequential and discrete time dynamics, respectively.
A heuristic derivation of a nonlinear diffusion equation is presented, and
the tracer diffusion coefficient is computed for arbitrary
$\phi(r)$ and different types of dynamics. 
\end{abstract}

\vspace{0.5cm}


\section{Introduction and outline}

In this paper we are concerned with systems of interacting particles
moving on the real line. The models of interest can be 
described as follows: Let $x_i \in \Bbb{R}$ denote the position of 
the $i$-th particle. In an elementary move particle $i$ jumps to 
the right to a position $x_i + \delta_i$ between $x_i$ and
$x_{i+1} > x_i$. 
In the absence of a lattice spacing, there are two natural
ways of setting the scale for the jump distance $\delta_i$:
It can be imposed externally through the choice of a fixed
probability density $f_i(\delta_i)$, in which case moves with
$\delta_i > x_{i+1} - x_i$ have to be rejected, or the scale
can be set by the  gap or ``headway''
\begin{equation}
\label{ui}
u_i = x_{i+1} - x_i
\end{equation}
in front of particle $i$ by letting $f_i$ depend on the
configuration ${\cal U} \equiv \{u_i \}_{i \in \Bbb{Z}}$ 
as
\begin{equation}
\label{pdf}
f_i(\delta_i \vert {\cal U}) = u_i^{-1} \phi(\delta_i/u_i),
\end{equation}
where $\phi(r)$ is a probability density with support on the unit interval. 
Equation (\ref{pdf}) implies that the jump length $\delta_i$ is a random
fraction $r$ of the headway $u_i$. 
The rate for the move is a function $\gamma(u_i)$ of the headway.
The moves are executed in continuous time (in which
case each particle is equipped with an exponential clock) or 
in discrete time; in the latter case the particle positions are updated
either in parallel, or sequentially by going through the system against the
direction of particle motion. A model is defined by specifying the functions
$f_i(\delta)$
and $\gamma(u)$ as well as the type of dynamics (continuous time,
parallel or sequential). 

Two equivalent representations of the dynamics will prove to be useful.
In terms of the headway
variables $u_i$ the particle configuration
may be visualized as a system of 
sticks located at the sites $i$ of the
integer lattice, $u_i$ being the length of stick $i$. In an elementary
move a fraction $\delta_i$ of stick $i$ is broken off and added to stick
$i-1$ \cite{timo96,rajesh}. Alternatively, the particle positions
$x_i(t)$ can be taken to define the height of a one-dimensional
interface above the point $i$. The asymmetric particle motion translates
into a growth process, and the fact that particles cannot pass each
other implies that the interface is a monotonically increasing
staircase ($x_{i+1} - x_i > 0$) at all times. We will refer to these
two viewpoints as the stick representation and
the interface representation, respectively.  

For continuous time dynamics, a jump length distribution of the 
type (\ref{pdf}) with $\phi$ uniform, 
and $\gamma(u) = u$ the model reduces to 
the Hammersley process discussed in \cite{aldous95}. In this case
the invariant distribution of particle positions is Poisson. 
Here we are interested in obtaining similar
results for other choices of  $f_i$ and 
$\gamma$, and other types of dynamics. 
Our motivation is mainly conceptual: While a wealth of results
\cite{liggett,spohn,kipnis99,liggett99,spohn}
are available for particle systems on the integer lattice 
such as the asymmetric simple exclusion process \cite{spitzer},
little is known analytically for the case of continuous
particle positions, although motion on the real line
appears naturally e.g. in applications to highway traffic 
\cite{krauss96,krauss97,krauss98}. 

An important simplifying feature of the asymmetric exclusion process
is the existence of stationary product measures. Here the analogous
desirable property is the product form
\begin{equation}
\label{product}
{\cal P}({\cal U}) = \prod_i P(u_i)
\end{equation}
for the stationary probability of a configuration
${\cal U}$ of particle headways. Therefore a
primary goal will be to find nontrivial examples of
asymmetric particle systems on $\Bbb{R}$ for which (\ref{product})
holds.

We provide an outline of the paper. In the next section we 
explore the conditions for a Poisson distribution of
particle positions (corresponding to an exponential distribution
of interparticle spacings in (\ref{product})) to be 
invariant 
for continuous time dynamics.
Our strategy is to consider a finite number $N$ of particles moving
on a ring of length $L$, and to demand that the stationary
measure gives the same weight to all allowed configuration;
this then implies a Poisson measure for $N,L \to \infty$
at fixed density $\rho = N/L$. 
Provided the jump rate $\gamma$ is independent
of the headway, we find that the Poisson measure is invariant
for {\em arbitrary} externally imposed (i.e., configuration
and particle
independent) jump length distributions $f(u)$.
On the other hand, if the jump length is scaled to the headway
as in (\ref{pdf}), the Poisson measure is stationary 
only for a one-parameter family of 
power law functions $\phi$ and $\gamma$, which have been identified
previously in the context of (symmetric) stick models \cite{feng96}.

Sections 3 and 4, which constitute the main part of the 
paper, are devoted to models with constant jump rate,
$\gamma \equiv 1$ independent of the headway, and jump length
distributions of the type (\ref{pdf}). 
In the interface representation these belong to the class  
of random average processes (RAP) studied by
Ferrari and Fontes \cite{ferrari98}:
The particle position $x_i'$ after the
move is an average
\begin{equation}
\label{rap}
x_i' = r x_i + (1-r) x_{i+1}
\end{equation}
of the previous positions $x_i$, $x_{i+1}$, with a random 
weight $r \in [0,1]$ drawn from the probability
density $\phi(r)$.  
We therefore refer
to these models as Asymmetric Random Average Processes (ARAP).
Discrete time ARAP's have been introduced previously
to model force fluctuations in random bead packs 
\cite{liu95,coppersmith96,claudin98}. In that context the headway $u_i(t)$
represents the (scaled) force supported by bead $i$ at depth
$t$ below the surface of a two-dimensional packing (see 
Section 3.2.1). 

In Section 3.1.1 we show, for the
case of continuous time dynamics, that the two-point correlation
function of particle headways $\langle u_i u_j \rangle$ factorizes
in the stationary state for any choice of $\phi(r)$, and obtain
the expression
\begin{equation}
\label{varu}
\langle u^2 \rangle - \langle u \rangle^2 = \frac{\mu_2}{\rho^2(\mu_1 - \mu_2)}
\end{equation}
for the stationary variance of headways in terms of 
the moments 
\begin{equation}
\label{mn}
\mu_n = \int_0^1 dr \; r^n \phi(r)
\end{equation}
of $\phi(r)$ and the particle density $\rho$. Similar results
for the discrete time models are derived in Section 3.2. 

More detailed information about the stationary headway
distribution can be obtained when $\phi(r)$ is the uniform
distribution on $[0,1]$. Assuming that the factorization property
of the two-point function implies pairwise independence of
the $u_i$, we 
derive and solve 
stationarity conditions for their moments, 
which show that the invariant density of 
headways (normalized to $\langle u_i \rangle = 1$) 
takes the form of a gamma distribution,
\begin{equation}
\label{Gamma}
P_\nu(u) = \frac{\nu^\nu}{\Gamma(\nu)} 
u^{\nu - 1} 
e^{-\nu u}
\end{equation}
where the parameter $\nu$ depends on the dynamics: 
For continuous time dynamics $\nu = 1/2$, while 
sequential and parallel dynamics yield $\nu = 1$ and
2, respectively. The result for parallel dynamics 
has been previously derived by Coppersmith {\em et al.}
\cite{coppersmith96}, who also gave an explicit proof of
the factorization property (\ref{product}).  
Equation (\ref{Gamma}) implies {\em bunching} of particles 
(enhanced density fluctuations compared to the 
Poisson measure) for continuous time 
dynamics ($\nu = 1/2$) and {\em antibunching} for parallel 
dynamics ($\nu = 2$).  
The associated
nontrivial particle-particle correlations are explicitly
computed in Section 3.3. 

Based on numerical simulations, we conjecture 
that the stationary single particle headway distribution is
exactly given by (\ref{Gamma}) for all three
types of dynamics. For continuous time dynamics
and a finite number of particles on a ring
the assumption of an invariant product measure is
examined in Section 3.1.2. Surprisingly,
we find that the product measure is {\em not} invariant
for the ARAP, although it {\em is} invariant for a related
symmetric stick model. This conclusion agrees with recent
results for the infinite system obtained by Rajesh and Majumdar
\cite{rajesh}. 

Section 4
is devoted to the large scale, long time behavior of the ARAP.
We derive a hydrodynamic equation of singular diffusion type, and
compute the tracer diffusion coefficient using a Langevin approach.
Since these results depend only on the stationary two-point function
of headways, they are valid for any choice of the jump
length distribution $\phi(r)$.  
Finally, some conclusions and 
open questions are formulated in Section 5.

\section{Models with invariant Poisson measures}
\label{Poisson}

\subsection{Constant invariant measure on the ring}

In this section we want to identify continuous time
dynamics which leave a  Poisson distribution of 
particle positions invariant. For this purpose we
first consider $N$ particles moving in continuous time
on a ring of length $L$, with density $\rho = N/L$.
Allowed headway configurations then satisfy the constraint
\begin{equation}
\label{constraint}
\sum_{i=1}^N u_i = L
\end{equation}
and the product measure (\ref{product}) is required
to hold on the set of configurations
defined by (\ref{constraint}). For an exponential
distribution $P(u) \sim e^{- \rho u}$ this implies
that all allowed headway configurations carry the same
weight $\Omega(N,L)^{-1}$, where
\begin{equation}
\label{omega}
\Omega(N,L) = \frac{L^{N-1}}{(N-1)!}
\end{equation}
denotes the volume of the set, i.e. the invariant
measure is {\em constant} on allowed configurations.
It is straightforward
to check that this implies Poisson measure
in the limit $N,L \to \infty$ at fixed density $\rho$.
For example, the distribution of a single headway on
the ring is given by
\begin{equation}
\label{PNL}
P_{N,L}(u) = \frac{\Omega(N-1,L-u)}{\Omega(N,L)} 
\to \rho e^{-\rho u}, \;\;\;\; N,L \to \infty
\end{equation}
while the joint distribution of the headways of
two neighboring particles is
\begin{equation}
\label{PNL2}
P_{N,L}(u_i, u_{i+1}) = 
\frac{\Omega(N-2,L-u_i - u_{i+1})}{\Omega(N,L)} 
\to \rho^2 e^{-\rho(u_i + u_{i+1})}, \;\;\;\; N,L \to \infty.
\end{equation}
A similar argument can be carried out for the probability distribution
of the particle positions on the ring. 

Invariance of the constant measure requires
the total transition rates for going into and out of any configuration
to balance. This yields the condition
\begin{equation}
\label{invariant}
\sum_{i=1}^N \int_0^{u_{i-1}} dw \; f_i(w \vert {\cal U}^{(i)}(w))  
\gamma(u_i+w) = 
\sum_{i=1}^N \int_0^{u_i} dw f_i(w \vert {\cal U}) \gamma(u_i)
\end{equation}
for any configuration ${\cal U}$, 
with the configuration ${\cal U}^{(i)}(w) = \{ u_j^{(i)}(w) 
\}_{j \in \Bbb{Z}}$ defined through 
\begin{equation}
\label{uprime} 
u_j^{(i)}(w) = \left\{ \begin{array}{l@{\quad:\quad}l}
u_i + w &  j = i \\ 
u_{i-1} - w & j = i-1 \\
u_j & {\rm else}
\end{array} \right.
\end{equation}
and periodic boundary conditions implied
in the summation over $i$. 
Note the upper integration limits, which ensure that 
particles cannot pass each other ($\delta_i \leq u_i$). 
Two examples of dynamics which satisfy (\ref{invariant}) will
be given in the following.

\subsection{Configuration-independent jump length distributions}

If the jump rate $\gamma$ is independent of headway, the
invariance condition (\ref{invariant}) is seen to hold for 
{\em any} jump length distribution $f(w)$ which is independent
of the configuration and of the particle label $i$. The stationary
speed $\overline v$ of particles at density $\rho$ is
then computed from
\begin{equation}
\label{speed}
\overline v = \gamma \rho \int_0^\infty du \; e^{-\rho u} 
\int_0^u dw \; w f(w),
\end{equation}
and the current follows from $j(\rho) = \rho {\overline v}(\rho)$. For example,
for jump lengths chosen uniformly in the unit interval one finds
\begin{equation}
\label{juni}
j(\rho) = \frac{\gamma}{\rho} [1 - (1 + \rho)e^{-\rho} ].
\end{equation}

It should be noted that in general the Poisson distribution is not 
the unique invariant measure. For example, if $f(w) = 0$ for
$w$ less than some minimum jump length $a$, then all configurations 
with $u_i < a$ for all $i$ are trivially invariant. Numerical simulations
indicate, however, that such ``absorbing'' states are typically
not reached,
even if the system is started very close to them. If $f(w) = 
\delta(w - 1)$ {\em and} the particles are started on the integer lattice,
the model reduces to the asymmetric exclusion process, which has
a geometric (rather than exponential) headway distribution.

\subsection{Scale-invariant models}

When the scale of the jumps is set by the headways, inserting
(\ref{pdf}) into (\ref{invariant}) 
and requiring the terms on both sides to cancel pairwise
yields
the following integral equation connecting
the functions $\phi$ and $\gamma$,
\begin{equation}
\label{integral}
\int_0^u dw \;\gamma(u'+w) \frac{\phi(w/(u'+w))}{u'+w} = \gamma(u),
\end{equation}
which should be true for all $u$, $u'$. Taking the derivative with 
respect to $u$ this becomes a differential equation for $\gamma$,
\begin{equation}
\label{diffgamma}
\frac{d \gamma}{du} = \frac{\gamma(v) \phi(u/v)}{v}
\end{equation}
with $v = u + u' \geq u$. Setting in particular $v = u$ we see that
$\gamma$ has to be a power law function,
\begin{equation}
\label{powergamma}
\gamma(u) = \gamma_0 u^{\alpha - 1},
\end{equation}
where $\gamma_0 > 0$ is a constant and $\alpha = 1 + \phi(1)$. 
Using (\ref{diffgamma}) the jump length distribution is then found 
to be also a power law,
\begin{equation}
\label{powerphi}
\phi(v) = (\alpha - 1) v^{\alpha - 2}.
\end{equation}
Normalizability of $\phi$ requires $\alpha > 1$. 

Equations (\ref{powergamma},\ref{powerphi}) define a one-parameter
family of models for which the Poisson distribution of positions
is invariant for an arbitrary number of particles $N$, the Hammersley
process being given by $\alpha = 2$. The corresponding
{\em symmetric} stick models, in which the broken-off piece is
distributed with equal probability to the left or right neighbor,
were considered by Feng {\em et al.} \cite{feng96}. Since $\gamma$
is a power law, these models are {\em scale invariant} in the 
sense that the average particle spacing $\langle u_i \rangle = 
1/\rho$ is the only
length scale in the problem. Therefore also the stationary
particle current $j$ is a power law function of the density. 
To compute it, we note that the average particle
speed is given by 
\begin{equation}
\label{speed2}
{\overline v} =  \langle \gamma(u_i) \delta_i \rangle = 
\rho \int_0^\infty du \, e^{-\rho u} \gamma(u) u \int_0^1 dv \; v \phi(v) = 
\gamma_0 (1 - 1/\alpha) \Gamma(\alpha + 1) \rho^{-\alpha}
\end{equation}
and therefore 
\begin{equation}
\label{j}
j(\rho) = \rho {\overline v} = \gamma_0 (1 - 1/\alpha) 
\Gamma(\alpha + 1) \rho^{1 - \alpha}.
\end{equation}

\section{Asymmetric Random Average Processes}
\label{ARAP}

The asymmetric random average process is a 
scale-invariant model characterized by a
jump length distribution of type (\ref{pdf}), 
and a constant jump rate
$\gamma \equiv \gamma_0 = 1$. 
The discussion is phrased most naturally in the stick representation,
and begins with the continuous time models.  

\subsection{Continuous time dynamics}

\subsubsection{Stationary headway correlations} 

Consider first 
the time evolution of the second moment $\langle u_i^2 \rangle$.
In a small
time interval $\Delta t$ two processes affecting $u_i$ may occur:
A random fraction $\delta_i$ of $u_i$ may be lost to $i-1$,
and a random fraction $\delta_{i+1}$ of $u_{i+1}$ may be gained from
$i+1$. Both processes occur with probability $\Delta t$. Thus
\begin{equation}
\label{Deltat}
\langle u_i^2 \rangle (t + \Delta t) = 
\Delta t [\langle (u_i - \delta_i)^2 \rangle + \langle (u_i + 
\delta_{i+1} )^2 \rangle ] + (1 - 2 \Delta t)
\langle u_i^2 \rangle (t).
\end{equation}
Stationarity then implies
\begin{equation}
\label{delta}
- 2 \langle \delta_i u_i \rangle + \langle \delta_i^2 \rangle + 
2 \langle \delta_{i+1} u_i \rangle + \langle \delta_{i+1}^2 \rangle = 0.
\end{equation}
Since $\delta_j = r_j u_j$ where $r_j$ is an independent random
variable with mean $\mu_1$ and second moment $\mu_2$,  
we have that
$\langle \delta_i u_i \rangle = \mu_1 \langle u_i^2 \rangle$,
$\langle \delta_i^2 \rangle = \langle \delta_{i+1}^2 \rangle =
\mu_2 \langle u_i^2 \rangle$ and
$\langle \delta_{i+1} u_i \rangle =
\mu_1 \langle u_i u_{i+1} \rangle$. Thus (\ref{delta}) becomes
\begin{equation}
\label{corr0}
(\mu_1 - \mu_2) \langle u_i^2 \rangle = \mu_1 \langle u_i u_{i+1} \rangle.
\end{equation}
Similarly for the general two-point function $C_k \equiv \langle u_i u_{i+k}
\rangle$ we obtain the stationarity condition
\begin{equation}
\label{Ck}
\mu_1 (C_{k+1} + C_{k-1} - 2 C_k ) = \mu_2 C_0(\delta_{k,1} + \delta_{k,-1} -
2 \delta_{k,0}),
\end{equation}
where translational invariance and symmetry ($C_k = C_{-k}$) of the
correlations has been used. Solving eq.(\ref{Ck}) starting from 
$k=0$ one finds
\begin{equation}
\label{Ck2}
C_k = [1 - (\mu_2/\mu_1)(1 - \delta_{k,0})] C_0.
\end{equation}
Imposing the boundary condition 
$\lim_{k \to \infty} C_k = 
\langle u_i \rangle^2 = 1/\rho^2$
for an infinite system of density $\rho$, eq.(\ref{Ck2}) then
shows that the two-point function factorizes for any $k \geq 1$ and
the variance of headways is given by (\ref{varu}). 

\subsubsection{Stationary headway distribution for uniform
$\phi(r)$}

We now specialize to the case when the distribution of scaled
jump lengths $\phi(r)$ is uniform in $[0,1]$, and assume that the
factorization property which was verified above for the two-point
function implies the pairwise independence of the
$u_i$. 
Then the stationarity condition for the $n$-th moment 
\begin{equation}
\label{statn}
\langle (u_i + \delta_{i+1})^n \rangle + \langle (u_i - \delta_i)^n \rangle = 
2 \langle u_i^n \rangle.
\end{equation}
yields (the index $i$ of $u_i$ is now dropped)
\begin{equation}
\label{sumn}
\sum_{k=0}^n \left( {n \atop k} \right) \frac{1}{k+1} 
[\langle u^{n-k} \rangle \langle u^k \rangle + (-1)^k \langle u^n \rangle ]
= 2 \langle u^n \rangle
\end{equation}
which can be rewritten as a recursion relation,
\begin{equation}
\label{recursion}
\langle u^n \rangle = \frac{n+1}{n-1} \sum_{k=1}^{n-1} 
\left( {n \atop k} \right) \frac{1}{k+1} 
\langle u^{n-k} \rangle \langle u^k \rangle.
\end{equation}
Evaluating this expression for $n=1,...,5$ we find that the relation
\begin{equation}
\label{moments}
\langle u^n \rangle = \left[\prod_{k=1}^n (2k-1) \right] 
\langle u \rangle^n
\end{equation}
appears to hold, which is characteristic of the gamma distribution
(\ref{Gamma}) with parameter $\nu = 1/2$. 

To prove it, we first 
insert (\ref{moments}) into (\ref{recursion}),
and obtain
\begin{equation}
\label{binom1}
\left( {2n \atop n} \right)=  \frac{n+1}{n-1} \sum_{k=1}^{n-1} 
\frac{1}{k+1} \left( {2k \atop k} \right) \left( {2(n-k) \atop n-k} \right).
\end{equation}
This can be verified using the binomial expansion 
\begin{equation}
\label{binom2}
\frac{1}{2} (1 - 4x)^{-1/2} = \frac{1}{2} + \sum_{k=1}^\infty 
\left( {2k-1} \atop {k-1} \right) x^k.
\end{equation}
Integrating with respect to $x$ we also have
\begin{equation}
\label{binom3}
-\frac{1}{4} (1 - 4x)^{1/2} = - \frac{1}{4} + \frac{x}{2} + 
\sum_{k=1}^\infty \frac{1}{k+1} \left( {2k-1} \atop {k-1} \right)
x^{k+1}. 
\end{equation}
Since the product of the left hand sides is a constant,
all coefficients of $x^m$ with $m > 0$ in the series obtained by multiplying
(\ref{binom2}) and (\ref{binom3}) must vanish. After rearranging terms
this is seen to imply (\ref{binom1}).

In fact the relation (\ref{moments}) was first guessed on the basis
of numerical simulations. Rather accurate numerical estimates
for the stationary 
moments of $u_i$ can be obtained by 
starting from an ordered initial condition
($u_i = 1$ for all $i$) and fitting the finite time 
data to the form
\begin{equation}
\label{finitetime}
\langle u^n \rangle(t) = A_n + B_n \; t^{-1/2}
\end{equation}
which is suggested by the fluctuation theory of Section 4.2 
(see eq.(\ref{G1})). 
The results shown in Table I 
strongly 
indicate that the stationary single particle 
headway distribution is exactly
given by the $\nu = 1/2$ gamma distribution.

To test the assumption of an invariant product measure
underlying the derivation of (\ref{recursion}), we proceed
as in Section 2 and consider
a finite number $N$ of particles on a ring.
The condition for the product measure (\ref{product}), 
restricted
to the set (\ref{constraint}) of allowed configurations,
to be invariant now 
reads
\begin{equation}
\label{invariantrap}
\sum_{i=1}^N \int_0^{u_{i-1}} 
\frac{dw}{u_{i} + w} \frac{P(u_i + w) P(u_{i-1} - w)}{
P(u_i) P(u_{i-1})} = 
\sum_{i=1}^N  \gamma(u_i) = N.
\end{equation}
Inserting the gamma distribution with parameter $\nu = 1/2$ 
(eq.(\ref{Gamma}))
and noting that
\begin{equation}
\label{integral2} 
\int_0^v dw (u+w)^{-3/2} (v - w)^{-1/2} = 
\frac{2 \sqrt{v/u}}{u + v}
\end{equation}
the condition (\ref{invariantrap}) becomes 
\begin{equation}
\label{invariant2}
\sum_{i=1}^N \frac{2 u_{i-1}}{u_i + u_{i-1}} = N, 
\end{equation}
with periodic boundary 
conditions, $u_{-1} = u_N$.
Equation (\ref{invariant2}) is satisfied for $N=2$, but not for 
general $N$. We conclude that the product measure (\ref{product})
is {\em not} invariant for
$N$ different from 2. It is however the exact invariant measure for
the {\em symmetric} stick process obtained by transferring the 
piece broken off stick $i$ to $i-1$ or $i+1$ with equal probability.
Indeed, in that case the left hand side of (\ref{invariant2}) becomes
\begin{equation}
\label{invariant3}
\sum_{i=1}^N \frac{u_{i-1}}{u_i + u_{i-1}} +
\frac{u_{i+1}}{u_i + u_{i+1}} = N. 
\end{equation}
While these arguments are restricted to finite systems, the conclusions
agree with calculations carried out for the infinite system by Rajesh and Majumdar
\cite{rajesh}. Specifically, they show that the product measure ansatz
for the continuous time ARAP breaks down at the level of 
three-point correlations,
but is exact for the symmetric stick model.

\subsection{Discrete time dynamics}

\subsubsection{Parallel update}
\label{parup}

A discrete time version of the ARAP is obtained by writing
\begin{equation}
\label{parallel}
u_i(t+1) = u_i(t) - \delta_i(t) + \delta_{i+1}(t)
\end{equation}
where $\delta_j = r_j u_j$ with independent random numbers
$r_j$ distributed according to the density
$\phi(r)$. This is closely related
to a model introduced by Coppersmith, Liu, Majumdar, Narayan and Witten
for the description of force fluctuations in bead packs
\cite{coppersmith96}. To see the connection, let $W(i,t)$ denote the
weight supported by bead $i$ in the $t$-th layer below the (free) surface
of the packing. The key assumption of the model is that the beads are
arranged on a regular lattice, and that each bead transfers its weight
to exactly $M$ beads in the layer below. The fraction $q_{ij}(t) \in [0,1]$ of
the weight of bead $i$ in layer $t$ which is transferred to bead $j$ in 
layer $t+1$ defines a matrix with random entries subject to the
constraint $\sum_{j} q_{ij}(t) = 1$. Assigning unit mass to each bead,
the weights evolve according to 
\begin{equation} 
\label{q-model}
W(j,t+1) = 1 + \sum_i q_{ij}(t) W(i,t).
\end{equation}
For large $t$ all weights increase linearly with $t$, which suggests
to introduce normalized variables $U(i,t) = W(i,t)/t$. 
Specializing to a two-dimensional lattice where the beads are labeled such
that bead $i$ is connected to beads $i$ and $i+1$ in the layer below,
we see that for $t \to \infty$ the evolution of the $U(i,t)$ reduces
to (\ref{parallel}) with the identification $q_{ii} = 1 - r_i$ and
$q_{i+1 i} = r_{i+1}$. In the context of beak packs $q_{ii}$ and 
$q_{i+1 i}$ should have the same distribution, and hence strict equivalence
between the two models holds only when $\phi(r)$ is symmetric around
$r=1/2$. 

Let us first show that the stationary two-point headway correlations  
factorize for any $\phi(r)$. Proceeding as above in Section 3.1.1, we
obtain the stationarity condition 
\begin{equation}
\label{Ckpar}
(\mu_1 - \mu_1^2) (C_{k+1} + C_{k-1} - 2 C_k ) = 
(\mu_2 - \mu_1^2) C_0(\delta_{k,1} + \delta_{k,-1} -
2 \delta_{k,0}),
\end{equation}
with the solution 
\begin{equation}
\label{Ckpar2}
C_k = [1 - (\mu_2-\mu_1^2)/(\mu_1-\mu_1^2)(1 - \delta_{k,0})] C_0.
\end{equation}
As in the continuous time case
this implies factorization
for $k \geq 1$ in the infinite system, with the stationary variance
of headways given by 
\begin{equation}
\label{varupar}
\langle u^2 \rangle - \langle u \rangle^2 = 
\frac{\mu_2 - \mu_1^2}{\rho^2(\mu_1 - \mu_2)}.
\end{equation}

For the case of a uniform distribution $\phi(r)$, Coppersmith {\em et al.}
\cite{coppersmith96} (see also \cite{majumdar99,rajesh})
have shown explicitly that the stationary measure takes
the product form (\ref{product}), with the headway distribution $P(u)$
given by the gamma distribution (\ref{Gamma}) with $\nu = 2$. The latter 
is easily derived along the lines of Section 3.1.2. 
Under the assumption of pairwise independence, the 
stationarity condition for general moments $\langle u_i^n \rangle$
now reads
\begin{equation}
\label{parrecursion}
\langle u^n \rangle = \frac{1}{(n-1)(n+2)} \sum_{k=1}^{n-1} 
\left( {n+2 \atop k+1} \right) 
\langle u^{n-k} \rangle \langle u^k \rangle.
\end{equation}
A straightforward computation shows that this 
is solved by the expression 
\begin{equation}
\label{parmom}
\langle u^n \rangle = 2^{-n} (n+1)! \langle u \rangle^n
\end{equation}
for the moments of the gamma distribution (\ref{Gamma}) with 
parameter $\nu = 2$. 

\subsubsection{Ordered sequential update}
\label{ordered}

In the context of traffic modeling \cite{evans97,raj98} 
it has been found useful
to implement a different kind of discrete time dynamics, in which 
the particles are moved one by one, in the order of their positions in 
the system. This {\em ordered sequential update} can proceed either in
the direction of particle motion (forward update) or against it (backward
update). For the ARAP it is easy to see that
the forward update is equivalent to the parallel dynamics discussed in 
Section 3.2.1, however the backward update is not. 

In the stick representation, backward sequential update implies that 
stick $i$ first receives a random fraction of stick $i+1$, placing it
in an intermediate state of length $u_i'$, and subsequently transfers
a random fraction $\delta_i'$ of $u_i'$ to stick $i-1$. It is important
to note that, at the time of transfer of mass to stick $i$, stick $i+1$
has already received mass from $i+2$ and thus the amount transferred from
$i+1$ to $i$ is a random fraction of $u_{i+1}' > u_{i+1}$. The dynamics
therefore proceeds in two steps,
\begin{equation}
\label{seq1}
u_{i}'(t) = u_i (t) + \delta_{i+1}'(t)
\end{equation}
\begin{equation}
\label{seq2}
u_{i}(t+1) = u_i' (t) - \delta_{i}'(t),
\end{equation}
where $\delta_j'$ is a random fraction of $u_j'$. 
Taking the average of both sides of (\ref{seq1}) or
(\ref{seq2}) the stationary mean of $u_i'$ is seen to be
\begin{equation}
\label{meanu'}
\langle u' \rangle = \frac{\langle u \rangle}{1 - \mu_1} = 
\frac{1}{\rho(1 - \mu_1)}. 
\end{equation}
Equation (\ref{seq2}) implies the relation
\begin{equation}
\label{C'}
C_k = [(1 - \mu_1)^2 + (\mu_2 - \mu_1^2)\delta_{k,0}] C_k'
\end{equation}
between the stationary two-point functions $C_k$ of $u_i$ and
$C_k'$ of $u_i'$. Using (\ref{seq1}) it is easy to show that
the stationarity condition for $C_k'$ is identical to the
condition (\ref{Ckpar}) obtained in the case of parallel update. 
Therefore also $C_k'$ factorizes in the infinite system, and
through (\ref{C'}) this property carries over to $C_k$. 
For the stationary variance of the backward sequential update 
model we find the expression
\begin{equation}
\label{varseq}
\langle u^2 \rangle - \langle u \rangle^2 = 
\frac{\mu_2 - \mu_1^2}{\rho^2(1-\mu_1)(\mu_1 - \mu_2)}.
\end{equation}

Turning to the stationary headway probability distribution $P(u)$, 
we again assume pairwise independence and note the
functional
equation
\begin{equation}
\label{functional}
P(u) = \int_0^1 dr \; r^{-1} \phi(1-r)  P'(u/r)
\end{equation}
relating $P(u)$ to the distribution $P'(u')$ of the intermediate
state headway. For uniform $\phi(r)$ the stationarity condition
for the $n$-th moment of $u_i'$ then reads
\begin{equation}
\label{statu'}
\langle (u_i')^n \rangle = \langle (u_i + \delta_{i+1}')^n 
\rangle = \sum_{k=0}^n \left( n \atop k \right)
\frac{1}{k+1} \langle (u_i')^k \rangle \langle u_i^{n-k} \rangle.
\end{equation}
Using the relation $\langle (u')^n \rangle = (n+1) \langle u^n \rangle$
obtained from (\ref{functional}) this reduces to 
\begin{equation}
\label{statuseq}
\langle u^n \rangle = \frac{1}{n+1} \sum_{k=0}^n 
\left( n \atop k \right) \langle u^k \rangle 
\langle u^{n-k} \rangle, 
\end{equation}
which is solved by setting $\langle u^n \rangle = n! \langle u \rangle^n$. 
We conclude that $P(u)$ is an exponential distribution (a gamma
distribution (\ref{Gamma}) with $\nu = 1$). This is confirmed by the
numerical data shown in Table I.

From (\ref{functional}) the distribution of the intermediate state headway
is found to be a $\nu=2$ gamma distribution with mean $2\langle u \rangle =
2/\rho$, 
\begin{equation}
\label{primedist}
P'(u) = \rho^2 u e^{-\rho u}.
\end{equation}
Given the equivalence between the intermediate state headway and
the headway for parallel update which we found on the level of the two-point
function, it is no surprise that (\ref{primedist}) is identical, up to
a scale factor, to the headway distribution $P(u)$ for parallel dynamics.

\subsection{Particle-particle correlations}

In this section we illustrate how the product measure
(\ref{product}) with the headway distribution (\ref{Gamma})
translates into nontrivial particle-particle correlations when
$\nu \neq 1$. 
For example, the probability density $g(x)$ for finding
a particle at $x$, conditioned on having a particle at the origin,
can be written as 
\begin{equation}
\label{g}
g(x) = \sum_{n=1}^\infty P_n(x),
\end{equation}
where $P_n(x)$ is the probability density for the $n$-th particle
to be at $x$ when the $0$-th is at the origin or, equivalently,
the probability that $\sum_{i=0}^{n-1} u_i = x$. The $P_n$ are obtained
iteratively from $P_1(x) = P(x)$ through the convolution
\begin{equation}
\label{convolve}
P_n(x) = \int_0^x dy \; P_{n-1}(y) P(x-y).
\end{equation}
Inserting the gamma distributions (\ref{Gamma}) with parameters
$\nu = 1/2$ and $\nu = 2$, one finds that 
\begin{equation}
\label{Pncont}
P_n(x) = \rho (\Gamma(n/2) 2^{n/2})^{-1} (\rho x)^{n/2 -1}
e^{- \rho x/2}
\end{equation}
for the continuous time case, and 
\begin{equation}
\label{Pnpar}
P_n(x) =  \frac{2^{2n} \rho}{(2n-1)!}(\rho x)^{2n-1}
e^{-2 \rho x}
\end{equation}
for parallel dynamics. 

In the parallel case the evaluation of the sum
(\ref{g}) is straightforward, and yields the expression
\begin{equation}
\label{gpar}
g(x) = \rho (1 - e^{-4 \rho x})
\end{equation}
for the correlation function, which explicitly displays the 
tendency of particles to avoid each other at distances short compared
to $1/\rho$. 

To compute (\ref{g}) with the $P_n$
given by (\ref{Pncont}), it is useful to write $g$ as the sum of
two contributions $g_{\rm even}$ and $g_{\rm odd}$ from even and
odd $n$, respectively. One finds that $g_{\rm even}(x) = \rho/2$ 
independent of $x$, while the odd part can be brought into the form
\begin{equation}
\label{godd}
g_{\rm odd}(x) = P_1(x) + \frac{\rho }{2 \sqrt{\pi}} 
e^{-\rho x/2} \sum_{m=1}^\infty \frac{(m-1)!}{(2m-1)!}
(\sqrt{2 \rho x})^{2m-1}.
\end{equation}
To sum the series we write $(m-1)! = \int_0^\infty dz \; z^{m-1}
e^{-z}$ and interchange the summation over $m$ with the integration
over $z$. This yields finally 
\begin{equation}
\label{gcont}
g(x) = \sqrt{\frac{\rho}{2 \pi x}} e^{-\rho x/2} +
\frac{\rho}{2}(1 + {\rm erf}{\sqrt{\rho x/2}})
\end{equation}
with the error function ${\rm erf}(z) = (2/\sqrt{\pi})
\int_0^z dt \; e^{-t^2}$.
For $x \to 0$ the correlation function is dominated by $P_1(x)$
and correspondingly diverges as $1/\sqrt{x}$, reflecting the
tendency of particles to bunch together in the continuous time case. 
For $x \to \infty$ $g(x)$ decays somewhat faster than exponentially,
as
\begin{equation}
\label{largex}
g(x) - \rho \approx \frac{\rho}{\sqrt{2 \pi} (\rho x)^3}
e^{-\rho x/2}.
\end{equation}

Alternatively the correlations between particles can be 
characterized through the variance $(\Delta N_L)^2$ of the
number of particles $N_L$ in an interval of size $L$. When
$L$ is small compared to the mean interparticle spacing 
$N_L$ is either 0 or 1, and $(\Delta N_L)^2 = \rho L$.
For $L \gg 1/\rho$ a central limit argument shows that
\begin{equation}
\label{DeltaN}
(\Delta N_L)^2 \approx \chi L
\end{equation}
where the ``compressibility'' $\chi$ (defined in analogy with 
equilibrium systems \cite{spohn}) is given by
\begin{equation}
\label{chi}
\chi(\rho) = \rho^3 (\langle u^2 \rangle - \langle u \rangle^2) = 
\rho/\nu,
\end{equation}
with the parameter $\nu$ of the headway distribution (\ref{Gamma}). 
Thus the slope of $(\Delta N_L)^2$ versus $L$ changes from
unity for $L \ll 1/\rho$ to $1/\nu$ for $L \gg 1/\rho$, reflecting
the increase (decrease) of particle number fluctuations for
continuous time (parallel) dynamics, respectively. 
The compressibility is related to the pair correlation function
(\ref{g}) through
\begin{equation}
\label{chig}
\chi = \rho ( 1 + \int_{- \infty}^\infty dx \; (g(x) - \rho)).
\end{equation}

\section{Large scale dynamics of the ARAP}
\label{hydro}

\subsection{Hydrodynamic equation}

The average particle speed ${\overline v}$
in the ARAP is inversely proportional
to the density, hence the current
$j = \rho {\overline v}$ is independent of $\rho$. 
The dynamics on the Euler scale $x \sim t$
is therefore trivial, and one expects a hydrodynamic equation of diffusion
type \cite{spohn}. A simple derivation will be given below.
Throughout this section we consider a general scaled jump length
distribution $\phi(r)$. 

\subsubsection{Continuous time dynamics}

In the continuous time case the ensemble averaged
particle positions $X_i \equiv \langle x_i \rangle$ 
evolve according
to the {\em linear} equations
\begin{equation}
\label{Xi}
\frac{d X_i}{dt} = \mu_1 (X_{i+1} - X_i).
\end{equation}
This problem has been studied previously in the context of crystal growth
\cite{wed97}, and the procedure can be directly applied to the present
context. 

To extract the long wavelength behavior,
we introduce a scaling parameter \cite{spohn,kipnis99} $\epsilon$ and a smooth
function $\xi(y,\tau)$ such that
\begin{equation}
\label{scale}
X_i(t) = \xi(\epsilon i, \epsilon t). 
\end{equation}
Inserting this into (\ref{Xi}) and expanding to second order in $\epsilon$
we obtain
\begin{equation}
\label{hydro1}
\mu_1^{-1}
\frac{\partial \xi}{\partial \tau} = \frac{\partial \xi}{
\partial y} + \frac{\epsilon}{2} \frac{\partial^2 \xi}{\partial y^2}. 
\end{equation}
In the scaling limit $\epsilon \to 0$ this becomes a first order
equation which describes simple translation to the left 
\cite{ferrari98}. 

Here we will however postpone to take the limit,
and first carry out a Lagrange
transformation \cite{rosenau95,wed97}, 
which relates the Lagrangian description in terms of
the particle positions $X_i(t)$ to the Eulerian evolution of the density
field. The local density $\rho$ near the position of particle $i$ is
estimated as $(X_{i+1} - X_i)^{-1}$, so using (\ref{scale}) we
have the relation
\begin{equation}
\label{rhoxi}
\rho(\xi(y,\tau), \tau) = \epsilon^{-1} (\partial \xi/\partial y)^{-1}. 
\end{equation}
Differentiating this equation with respect to $\tau$ and using
the evolution equation (\ref{hydro1}) for $\xi(y,\tau)$ one obtains,
after some algebra,
\begin{equation}
\label{hydro2}
\frac{\partial \rho}{\partial t} = 
\epsilon \frac{\partial \rho}{\partial \tau} = 
\frac{\partial}{\partial x}
\left(\frac{\mu_1}{2 \rho^2} \right) \frac{\partial \rho}{\partial x}.
\end{equation}
The scaling factor $\epsilon$ cancels, and 
the {\em collective} diffusion coefficient is identified to be
\begin{equation}
\label{Dcoll}
D_{\rm c}(\rho) = \frac{\mu_1}{2 \rho^2}.
\end{equation}
The $\rho^{-2}$-dependence is dictated by scale invariance: The typical
jump length in a region of density $\rho$ is ${\overline \delta} =
\mu_1/\rho$, and $D_{\rm c} \sim \gamma {\overline \delta}^2 \sim \rho^{-2}$.

\subsubsection{Discrete time dynamics}

For discrete parallel update  eq.(\ref{Xi}) is replaced by
\begin{equation}
\label{Xidis}
X_i(t+1) - X_i(t) = \mu_1 [X_{i+1}(t) - X_i(t)].
\end{equation}
In the scaling limit $\epsilon \to 0$ this results in the same 
coarse grained evolution equation (\ref{hydro1}), and thus also
the nonlinear diffusion equation (\ref{hydro2}) is the same as in 
the continuous time case. 

In the case of ordered sequential update
one has to take into account that the new position of particle $i$
is a random average of its old position and the {\em new} position
of particle $i+1$, hence
\begin{equation}
\label{Xiseq}
X_i(t+1) - X_i(t) = \mu_1 [X_{i+1}(t+1) - X_i(t)].
\end{equation}
Making the ansatz $X_i(t) = i/\rho + {\overline v} t$, we see that
the average particle speed is 
\begin{equation}
\label{speedseq}
{\overline v} = \frac{\mu_1}{\rho(1-\mu_1)} > \frac{\mu_1}{\rho}.
\end{equation}
The speedup compared to continuous time and 
parallel dynamics is due to 
the decrease of the local density near the update site, see
\cite{raj98} for a discussion of similar effects in the asymmetric
exclusion process. For the derivation of the hydrodynamic equation
it is useful to incorporate the expected diffusive scaling from the
outset and replace (\ref{scale}) by   
\begin{equation}
\label{scaleseq}
X_i(t) = \xi(\epsilon i, \epsilon^2 t). 
\end{equation}
The expansion of (\ref{Xiseq}) to second order in $\epsilon$ then yields 
\begin{equation}
\label{hydro1seq}
\left(
\frac{1-\mu_1}{\mu_1}
\right)
\frac{\partial \xi}{\partial \tau} = \epsilon^{-1} \frac{\partial \xi}{
\partial y} + \frac{1}{2} \frac{\partial^2 \xi}{\partial y^2}. 
\end{equation}
As before, the drift term disappears under the Lagrange transformation
based on the relation (\ref{rhoxi}), and one obtains
\begin{equation}
\label{hydro2seq}
\frac{\partial \rho}{\partial t} = 
\epsilon^2 \frac{\partial \rho}{\partial \tau} = 
\frac{\partial}{\partial x}
\left(\frac{\mu_1}{2(1-\mu_1)\rho^2} \right) \frac{\partial \rho}{\partial x}.
\end{equation}
As far
as the hydrodynamics is concerned, the different types of dynamics
are seen to be equivalent up to a rescaling of time.

\subsection{Tracer diffusion}

Hydrodynamic equations of diffusion type are usually associated
with symmetric (unbiased) particle systems \cite{spohn}. In one
dimension the tracer diffusion coefficient in such systems 
typically vanishes, and the mean square displacement of a tagged
particle grows subdiffusively as $t^{1/2}$ \cite{arratia83,vB83}. By contrast,  
the biased random average process shows normal tracer diffusion when started 
from a random initial condition and subdiffusive behavior when 
the initial configuration is ordered \cite{ferrari98}. Here we
provide a compact derivation of the two cases and compute the
coefficient of the asymptotic law for different types of dynamics.

\subsubsection{Langevin approach for continuous time dynamics}

We start the system in an initial condition without long wavelength
fluctuations, such as $x_i(0) = i/\rho$, $i \in \Bbb{Z}$, and
denote the positional fluctuation of particle $i$ by
\begin{equation}
\label{zetai}
\zeta_i(t) = x_i(t) - \langle x_i \rangle = x_i(t) - x_i(0) - 
{\overline v} t.
\end{equation}
For the purpose of extracting the long time behavior of fluctuations,
a Langevin approximation \cite{vankampen92} 
to the dynamics of $\zeta_i$ is sufficient.
Thus we add a phenomenological noise term $\eta_i(t)$ to the 
linear equation (\ref{Xi}),
\begin{equation}
\label{Langevin}
\frac{d \zeta_i}{dt} = \mu_1 (\zeta_{i+1} - \zeta_i) + 
\eta_i.
\end{equation}
The noise is taken Gaussian with zero mean and covariance
\begin{equation}
\label{eta}
\langle \eta_i(t) \eta_j(t') \rangle = \sigma \delta_{ij} \delta(t - t').
\end{equation}
The noise strength $\sigma$ will eventually be matched to the variance 
of particle headways.

Equation (\ref{Langevin}) is solved by introducing the Fourier transformed
fluctuations
\begin{equation}
\label{ft}
\hat \zeta(q,t) = \sum_{n \in \Bbb{Z}} e^{i q n} \zeta_n(t)
\end{equation}
with wave numbers $q$ in the first Brillouin zone $[-\pi, \pi]$, and
the corresponding Fourier transformed noise 
\begin{equation}
\label{ftnoise}
\hat \eta(q,t) = \sum_{n \in \Bbb{Z}} e^{i q n} \eta_n(t)
\end{equation}
with covariance
\begin{equation}
\label{etaq}
\langle \hat \eta(q,t) \hat \eta(q',t') \rangle = 
2 \pi \sigma \delta(q + q') \delta(t - t').
\end{equation}
The most general 
quantity of interest is the variance of the displacement
between particle $i$ at time $t$ and particle $j$ at time $t'$.
By translational invariance this depends only on $n = i - j$ and
is given by the correlation function 
\begin{equation}
\label{Gn}
G_n(t, t') = \langle (\zeta_0(t) - \zeta_n(t'))^2 \rangle.
\end{equation}
Inserting (\ref{ft}) into (\ref{Langevin}), solving the equation
for $\hat \zeta(q,t)$ and averaging over the noise according to 
(\ref{etaq}) one arrives at the expression
$$
G_n(t,t') = 
$$
\begin{equation}
\label{Gn2}
\frac{\sigma}{2 \pi} \int_{0}^\pi 
\frac{dq}{\omega(q)}(2 - e^{-2 \omega t} - e^{-2 \omega t'} -
2 \cos[qn - \mu(q) T] (e^{-\omega \vert T \vert} - e^{-\omega T'}))
\end{equation}
with $\omega(q) = \mu_1(1 - \cos(q))$, $\mu(q) = \mu_1 \sin(q)$, 
$T = t' - t$ and $T' = t' + t$.

The evaluation is straightforward in the relevant limiting cases.
Consider first the 
variance of the headways at time $t = t'$. For large $t$ (\ref{Gn2}) yields
\begin{equation}
\label{G1}
G_1(t,t) \approx \frac{\sigma}{\mu_1} \left(
1 - \frac{1}{2 \sqrt{\pi \mu_1 t}}
\right).
\end{equation}
This allows us to identify
the noise strength $\sigma$ as
\begin{equation}
\label{sigmau}
\sigma = \mu_1 (\langle u^2 \rangle - \langle u \rangle^2),
\end{equation}
and explicitly demonstrates the $1/\sqrt{t}$-approach to the stationary
headway distribution alluded to in (\ref{finitetime}).

Next we focus on the dynamics of a single particle and
set $n=0$ in (\ref{Gn2}). If we fix the time increment $T = t' - t$
and let both $t$ and $t' \to \infty$, $G_0$ represents the mean square
displacement of a particle in the stationary regime. Evaluation of 
(\ref{Gn2}) gives $G_0(t,t') \approx \sigma \vert T \vert$, which
shows that $\sigma$ is precisely the tracer diffusion coefficient
$D_{\rm tr}$. Combining this with (\ref{sigmau}) and (\ref{varu}) 
we obtain
\begin{equation}
\label{tracer}
D_{\rm tr} = \mu_1 (\langle u^2 \rangle - \langle u \rangle^2) = 
\frac{\mu_1 \mu_2}{\rho^2(\mu_1 - \mu_2)}. 
\end{equation}
In fact the first relation in 
(\ref{tracer}) is easy to understand. 
The linear equation (\ref{hydro1}) shows that fluctuations in the particle
positions drift backwards
in ``label space'' $y = \epsilon i$. 
This translates the stationary distance fluctuations into
temporal fluctuations, with a conversion factor given by the drift 
speed $\mu_1$. 
As was mentioned already,
the existence of a nonvanishing tracer diffusion coefficient
for models with a hydrodynamic equation of diffusion type is unusual
in one dimension, since generically such an equation implies
symmetric particle jumps, in which case the tracer particle
displacement grows only subdiffusively  
due to the single file constraint \cite{arratia83,vB83}. 
Here $D_{\rm tr}$ is nonzero because the particles move, at speed
${\overline v}$, relative to the (stationary) density fluctuations.
A rigorous derivation of (\ref{tracer}) has recently been presented
by Sch\"utz \cite{schuetz99}.

Since the hydrodynamic equations in the two cases are identical,
the argument leading to first relation in (\ref{tracer}) 
carries over directly to discrete parallel update, and using
(\ref{varupar}) we conclude that the tracer diffusion coefficient
in this case is given by 
\begin{equation}
\label{tracerpar}
D_{\rm tr}^{\rm par} = \frac{\mu_1(\mu_2 - \mu_1^2)}{\rho^2(\mu_1 - \mu_2)}. 
\end{equation}
Similarly the expression 
\begin{equation}
\label{tracerseq}
D_{\rm tr}^{\rm seq} = 
\frac{\mu_1(\mu_2 - \mu_1^2)}{\rho^2(1-\mu_1)^2(\mu_1 - \mu_2)}. 
\end{equation}
is obtained for the backward sequential case 
by combining eqs.(\ref{varseq}) and (\ref{speedseq}).
Both (\ref{tracerpar}) and (\ref{tracerseq}) have been verified
numerically for the case of uniform $\phi(r)$. 

Subdiffusive behavior is found in the 
mean square displacement of a particle
starting from an initial configuration without long wavelength
disorder \cite{ferrari98}. This is given by (\ref{Gn2}) with
$n = t'= 0$. For large $t$ one obtains
\begin{equation}
\label{subdiff}
\langle \zeta_0^2 (t) \rangle = G_0(t,0) \approx \sigma 
\sqrt{\frac{t}{\pi \mu_1}}  
= 
\frac{\mu_2}{\rho^2(\mu_1 - \mu_2)} \sqrt{\frac{\mu_1 t}{\pi}}. 
\end{equation}
Using (\ref{chi}) and (\ref{Dcoll}) this is seen to agree with the
expression 
\begin{equation}
\label{subdiff2}
\langle \zeta_0^2 (t) \rangle = 
\sqrt{2/\pi} (\chi/\rho^2) \sqrt{D_{\rm c} t}
\end{equation}
derived from hydrodynamic arguments \cite{vB83}. 

\subsubsection{The independent jump approximation}

For the totally asymmetric simple exclusion process it is known 
\cite{spitzer,arratia83,ferrari96} 
that the motion of a tagged particle in the 
stationary state follows a Poisson process, and therefore the 
tracer diffusion coefficient is simply equal to the mean speed
$1 - \rho$. Here we show that the expressions (\ref{tracer} -
\ref{tracerseq}) for the ARAP are consistent with a similar
independent jump picture. 

Consider first the case of discrete time dynamics, where 
the random choice of the jump length $\delta_i$ is the only
source of disorder, and therefore the tracer diffusion coefficient
for independent jumps is equal to the variance of $\delta_i$. 
For parallel update $\delta_i$ is a 
uniform random fraction of the particle headway $u_i$, hence 
$
\langle \delta^2 \rangle - \langle \delta \rangle^2 = \mu_2
\langle u^2 \rangle - \mu_1^2/\rho^2$, which is easily checked to 
coincide with (\ref{tracerpar}). For the backward sequential case
$\delta_i$ is a random fraction of the intermediate state headway 
$u_i'$. Therefore, using eqs. (\ref{C'}), (\ref{meanu'}) and (\ref{varseq}),
\begin{equation}
\label{deltaseq}
\langle \delta^2 \rangle - \langle \delta \rangle^2 = 
\mu_2 \langle (u')^2 \rangle - \mu_1^2 \langle u' \rangle^2 = 
\frac{1}{(1-\mu_1)^2} \left(\frac{\mu_2 \langle u^2 \rangle}{
1 - 2 \mu_1 + \mu_2} - \frac{\mu_1^2}{\rho^2} \right),
\end{equation}
which is also found to agree with (\ref{tracerseq}). 

In the continuous time case the random timing of jumps introduces an 
additional source of disorder. It is natural to assume, in analogy
with the asymmetric exclusion process, that the 
jumps occur according to a Poisson process. In the independent jump
approximation the particle displacement $\Delta x$ in time $t$ is then
given by 
\begin{equation}
\label{deltax}
\Delta x (t) = \sum_{l=1}^{n(t)} \delta^{(l)}
\end{equation}
where $n(t)$ is a Poisson random variable with mean $t$ and the jump lengths
$\delta^{(l)}$ are independent random fractions of the 
(independent, random) particle headways.
It is straightforward to show that the variance of $\Delta x$
is 
\begin{equation}
\label{deltaxvar}
\langle (\Delta x)^2 \rangle - 
\langle \Delta x \rangle^2 = \langle \delta^2 \rangle t,
\end{equation}
thus in this case the independent jump approximation to 
$D_{\rm tr}$ is
$
\langle \delta^2 \rangle = \mu_2 \langle u^2 \rangle $
in agreement with (\ref{tracer}).

\section{Summary and outlook}
\label{open}

We have presented results for two classes of particle systems on 
$\Bbb{R}$. The models considered in Section 2. have 
Poisson invariant measures and nonlinear current-density relations 
(see eqs.(\ref{juni}, \ref{j})). Time-dependent fluctuations in these
models are therefore expected \cite{bks85,ks91}
to be governed by the noisy Burgers 
(or Kardar-Parisi-Zhang \cite{kpz86}) equation, which is not
amenable to simple analysis. By contrast, the asymmetric random
average processes introduced in 
Section 3. have nontrivial invariant measures, but
the linearity of the jump rules allows for a detailed
study of dynamic properties (Section 4.). 

A central result for the ARAP is the dependence of the 
headway distribution (\ref{Gamma}) 
on the type of dynamics. 
The idea that parallel update reduces density fluctuations
is familiar from earlier work on the asymmetric exclusion process
and related models for traffic flow, however in that case
the ordered sequential update produces the same (Bernoulli) invariant
measure as the continuous time process \cite{raj98}. 

Our study suggests that the invariant measure of the continuous time
ARAP displays an unusual combination of features: The two-point headway
correlations factorize, the single particle 
headway distribution appears to be exactly
given by the expression (\ref{Gamma})
derived under the assumption of pairwise independence,
but nevertheless the product measure (\ref{product}) 
is {\em not} invariant. Rajesh and
Majumdar have found the same features in a larger class of models which
interpolate between continuous time and parallel update \cite{rajesh}. 
It would be most interesting to find a simple ``deformation'' of the product
measure which explains this behavior. The status of the product measure
assumption for the ordered sequential update also remains to be clarified.
The considerations of Section 3.2.2 indicate that it might be possible
to exactly reduce this case to that of parallel update, for which the
product measure is known to be invariant \cite{coppersmith96}.

Another interesting direction for future work is the introduction
of quenched random inhomogeneities. 
In asymmetric 
exclusion models it is possible to find invariant product measures
also in the presence of random jump rates associated with particles
\cite{benjamini96,kf,evans96,evans97}. 
For the continuous time ARAP with jump rates $\gamma_i$ depending on the
particle label $i$ (the position $i$ in the stick representation) 
preliminary numerical
simulations indicate that the product measures discussed above do not
persist. It is possible to write down a closed set of linear equations
for the two-point function $\langle u_i u_j \rangle$ which depends on the
disorder configuration $\{ \gamma_i \}$ and which should yield insight
into the emergence and nature of correlations. 
Here we merely remark that, since the 
mean speed of particle $i$ is
$\gamma_i \langle u_i \rangle$, stationarity implies 
$
\langle u_i \rangle = C/\gamma_i
$
where the constant $C$ is fixed by the average headway. If the distribution
of jump rates is chosen such that $\langle 1/\gamma_i \rangle$ exists,
$C \to \langle 1/\gamma_i \rangle^{-1}$ in the limit 
of infinite system size, and all headways have a finite mean.
Otherwise (e.g. for a uniform distribution of jump rates) 
arbitrarily large headways will open in front of the slowest particles, similar to the
low density phase of asymmetric exclusion models with particlewise
disorder \cite{kf,evans96,evans97}.

\vspace{0.5cm}

{\bf Acknowledgements.} We are much indebted to 
Bernard Derrida, Pablo Ferrari,
Herve Guiol, Satya Majumdar, Gunter Sch\"utz and Timo 
Sepp\"al\"ainen for useful discussions
and remarks. This work was supported by DAAD and CAPES within the PROBRAL programme. J.K. acknowledges the hospitality of IME/USP, 
S\~{a}o Paulo, and the Erwin Schr\"odinger Institute
for Mathematical Physics, Vienna, where part of the paper was written.
J.G. acknowledges the hospitality of Universit\"at GH
Essen during the early stages of the project.

\vspace{2.cm}

\begin{center}
\begin{tabular}{|c||c|c|c|c|} \hline
\hfil Dynamics: &  $\langle u^2 \rangle$ & 
$\langle u^3 \rangle$ & $\langle u^4 \rangle$ & $\langle u^5 \rangle $
\\ \hline
Continuous &  $2.998 \pm 1$ & $15.02 \pm 1$ & $105.3 \pm 2$ & $947 \pm 3$ \\ 
  &  (3) & (15) &(105) & (945)   \\ \hline
Sequential  & $1.9997 \pm 2$ & $5.995 \pm 2$ & $23.93 \pm 2$ & $119.2 \pm 2$  \\ 
  &  (2) &  (6) & (24) & (120)  \\ \hline
Parallel  & $1.4998 \pm 1$ & $2.996 \pm 1$ &  $7.466 \pm 3$ & $22.26 \pm 2$ \\
  &  (3/2) & (3) & (15/2) & (45/2)  \\ \hline
\end{tabular}
\end{center}

\noindent
{\bf Table I.} The Table contains numerical estimates of the first few
moments of the stationary headway distribution for the ARAP with
uniform $\phi(r)$ and different kinds of update. The data were obtained
from simulations of systems of $2 \times 10^5$ particles which
were started from an ordered initial
condition, $u_i = 1$ for all $i$, 
and allowed to evolve for $10^4$ time steps. To extrapolate to 
$t \to \infty$, each run was fitted to eq.(\ref{finitetime}), and
the errors were estimated by taking an average over 10 runs
(errors refer to the last digit shown). 
The numbers in parentheses are the conjectured values of the
moments; for the case of parallel update these are known to be exact
\cite{coppersmith96}. The remaining discrepancies are in fact
largest for parallel update, and can probably be attributed to residual finite
time effects.

\end{document}